\documentclass{pnastwo}

\usepackage[dvips]{graphicx}
\usepackage{pnastwof}
\usepackage{amssymb,amsfonts,amsmath}

\newcommand{\nhppp}{non-homogeneous Poisson process}
\newcommand{\cnhppp}{cascading non-homogeneous Poisson process}

\newcommand{\SI}{\textit{SI}}
\newcommand{\suppinfo}[1]{see \SI~#1}
\newcommand{\methods}{see \textit{Methods}}
\newcommand{\figref}[1]{Fig.~#1}

\newcommand{\eqnref}[1]{Eq.~[#1]}

\newcommand{\secref}[1]{Sec.~#1}

\newcommand{\dataref}{S1}
\newcommand{\teststatref}{S2}
\newcommand{\mchtref}{S3}
\newcommand{\tplref}{S4}
\newcommand{\simannealref}{S5}

\newcommand{\simannealfigref}{S5}

\newcommand{\cnhpppfigref}{S7}

\newcommand{\cnhpppnrejected}{one}
\newcommand{\tplnrejected}{344}
\newcommand{\nusers}{394}

\begin{document}




\title{A Poissonian explanation for heavy tails in e-mail communication}

\author{
  R.~Dean Malmgren\affil{1}{Department of Chemical and
    Biological Engineering, Northwestern University, Evanston, IL
    60208, USA.}, 
  Daniel B.~Stouffer\affil{1}{}, 
  Adilson E.~Motter\affil{2}{Department of Physics and Astronomy,
    Northwestern University, Evanston, IL 60208,
    USA}\affil{3}{Northwestern Institute on Complex Systems, Evanston,
    IL 60208, USA}, \and 
  Lu\'is A.~N.~Amaral\affil{1}{}\affil{3}{}\thanks{To whom
    correspondence should be addressed. E-mail:
    amaral@northwestern.edu} 
}

\contributor{Submitted to Proceedings of the National Academy of
  Sciences of the United States of America}

\maketitle

\begin{article}

\begin{abstract}
Patterns of deliberate human activity and behavior are of utmost
importance in areas as diverse as disease spread, resource allocation,
and emergency response.  Because of its widespread availability and
use, e-mail correspondence provides an attractive proxy for studying
human activity.
Recently, it was reported that the probability density for the
inter-event time $\tau$ between consecutively sent e-mails decays
asymptotically as $\tau^{-\alpha}$, with $\alpha \approx 1$.
The slower than exponential decay of the inter-event time distribution
suggests that deliberate human activity is inherently non-Poissonian.
Here, we demonstrate that the approximate power-law scaling of the
inter-event time distribution is a consequence of circadian and weekly
cycles of human activity.  We propose a \cnhppp~which explicitly
integrates these periodic patterns in activity with an individual's
tendency to continue participating in an activity.  Using standard
statistical techniques, we show that our model is consistent
with the empirical data.
Our findings may also provide insight into the origins of heavy-tailed
distributions in other complex systems.
\end{abstract}



\dropcap{T}he 
%
analysis of social and economic data has a long and illustrious
history~\cite{smith86,pareto06,zipf49}. Despite their idiosyncratic
complexity, a number of striking statistical regularities are known to
describe individual and societal human
behavior~\cite{stanley96,huberman98,plerou99,amaral01a}.
These regularities are of enormous practical importance because they
provide insight into how individual behaviors influence social and
economic outcomes.  Indeed, much of the current research on complex
systems aims to quantify the impact of individual agents on the
organization and dynamics of the system as a
whole~\cite{newman03,castellano07}.  Before we can predict how
individuals affect, for example, the organization of systems, it is
paramount to understand the behavior of the individual agents.

The current availability of digital records has made it much easier
for researchers to quantitatively investigate various aspects of human
behavior~\cite{johansen00,johansen01,chessa04,johansen04,barabasi05,oliveira05,stouffer06,vazquez06,vazquez06a,dezso06,nakamura07,candia08}.
In particular, e-mail communication records are attracting much
attention as a proxy for quantifying deliberate human behavior due to
the omnipresence of e-mail communication and availability of e-mail
records~\cite{johansen04,barabasi05,vazquez06a,stouffer06}.  The data,
however, does not provide a detailed record of all of the activities
in which each individual participates; we do not know, for instance,
when an individual is sleeping, eating, walking, or even browsing the
web.  The resulting uncertainty in deliberate human activity thus
poses a fundamental challenge to quantifying and modeling human
behavior.

Researchers commonly account for uncertainty or lack of information
through stochastic models.  One of the simplest stochastic models for
human activity is a point process in which independent events occur at
a constant rate $\rho$.  Such processes are referred to as
\emph{homogeneous} Poisson processes, and they are used to describe a
large class of phenomena, including some aspects of human
activity~\cite{daley88}.  Homogeneous Poisson processes have two
well-known statistical properties: the time between consecutive
events, the inter-event time $\tau$, follows an exponential
distribution, $p(\tau) = \rho e^{-\rho\tau}$, and the number of events
$N_T$ during a time interval of duration $T$ time units follows a
Poisson distribution with mean $\rho T$.

Several recent studies of deliberate human activity, including e-mail
correspondence, have focused on the former property.
These studies have reported that the empirical distribution of
inter-event times decays asymptotically as a power-law, $p(\tau)
\propto \tau^{-\alpha}$, with exponent $\alpha \approx
1$~\cite{eckmann04,johansen04,barabasi05,vazquez06a}.  Other studies
have identified a similar power-law scaling in the inter-event time
distribution of many other facets of human behavior, such as file
downloads~\cite{johansen00,johansen01,chessa04}, letter
correspondence~\cite{oliveira05,vazquez06,vazquez06a}, library
usage~\cite{vazquez06}, broker trades~\cite{vazquez06}, web
browsing~\cite{vazquez06,dezso06}, human locomotor
activity~\cite{nakamura07}, and telephone
communication~\cite{candia08}.  These observations are in stark
contrast with the predictions of a homogeneous Poisson process,
suggesting that a more suitable null model with which to compare
mechanistic models of human activity is a truncated power-law model
with scaling exponent $\alpha = 1$.\footnote{For simplicity we use a
truncated power-law with an exponent of $\alpha=1$ as our null model.
Similar conclusions are reached when the power-law scaling exponent is
fit to the data or when other heavy-tailed null models
(\emph{e.g.} log-normal or log-uniform
distributions~\cite{stouffer06}) are considered.}

The heavy-tailed nature of the distribution of inter-event times prompts us to
search for the mechanisms responsible for its emergence.  Two main classes of
mechanisms can be considered:
(\emph{i}) human behavior is primarily
driven by rational decision making, which introduces correlations in activity
thereby giving rise to heavy-tails;\footnote{If humans make decisions based on
  their own previous memories, then we might expect that humans are heavily
  influenced by recent events.  That is, the probability $\rho\,dt$ that an
  event will happen in a time interval $dt$ is not constant, but is instead a
  decreasing function of the time elapsed since the last
  event~\cite{vazquez06a}.}
(\emph{ii}) human behavior is primarily driven by external factors such as
circadian and weekly cycles, which introduces a set of distinct characteristic
time scales thereby giving rise to heavy-tails.\footnote{This interpretation
  does not rely on highly competent human behavior and allows for the
  possibility that human activity, and hence the time dependence of $\rho$, is
  modulated by instinct, the environment, or social stimuli.}
While the former interpretation has been shown to give rise to a truncated
power-law distribution of inter-event times, the latter has been rejected by
some authors~\cite{vazquez06}.
Indeed, even though Hidalgo~\cite{hidalgo06} investigated a model with seasonal
changes in activity rates that is able to generate data with an approximate
power-law decay in the distribution of inter-event times with exponents $\alpha
\approx 2$ or $\alpha \approx 1$, the $\alpha \approx 1$ case requires a
specific relationship between the rates of activity $\rho_i$ and the
corresponding duration of the seasons $T_i$ over which each rate holds.  It has
therefore been argued that seasonality alone can only robustly give rise to
heavy-tailed inter-event time distributions with exponent $\alpha \approx
2$~\cite{vazquez06}.

Here, we demonstrate that the distribution of inter-event times in
e-mail correspondence patterns display systematic deviations from the
truncated power-law null model due to circadian and weekly patterns of
activity.  We subsequently propose a mechanistic model that
incorporates these observed cycles, and a novel simulated annealing
procedure to nonparametrically estimate its parameters.  We then use Monte Carlo
hypothesis testing to demonstrate that the predictions of our model
are consistent with the observed heavy-tailed inter-event time
distribution.  Finally, we discuss the implications of our findings on
modeling human activity patterns and, more generally, complex systems.

\section{Empirical patterns}

We study a database of e-mail records for 3,188 e-mail accounts at
a European university over an 83-day period~\cite{eckmann04}.  Each
record comprises a sender identifier, a recipient identifier, the size
of the e-mail, and a time stamp with a precision of one second.  We
preprocess the data set and identify a set of \nusers~accounts which
provide enough data to quantify human activity and which are likely
neither spammers nor listservs (\suppinfo{\secref{\dataref}}).

\begin{figure}
\center{\includegraphics*[width=1.0\columnwidth]{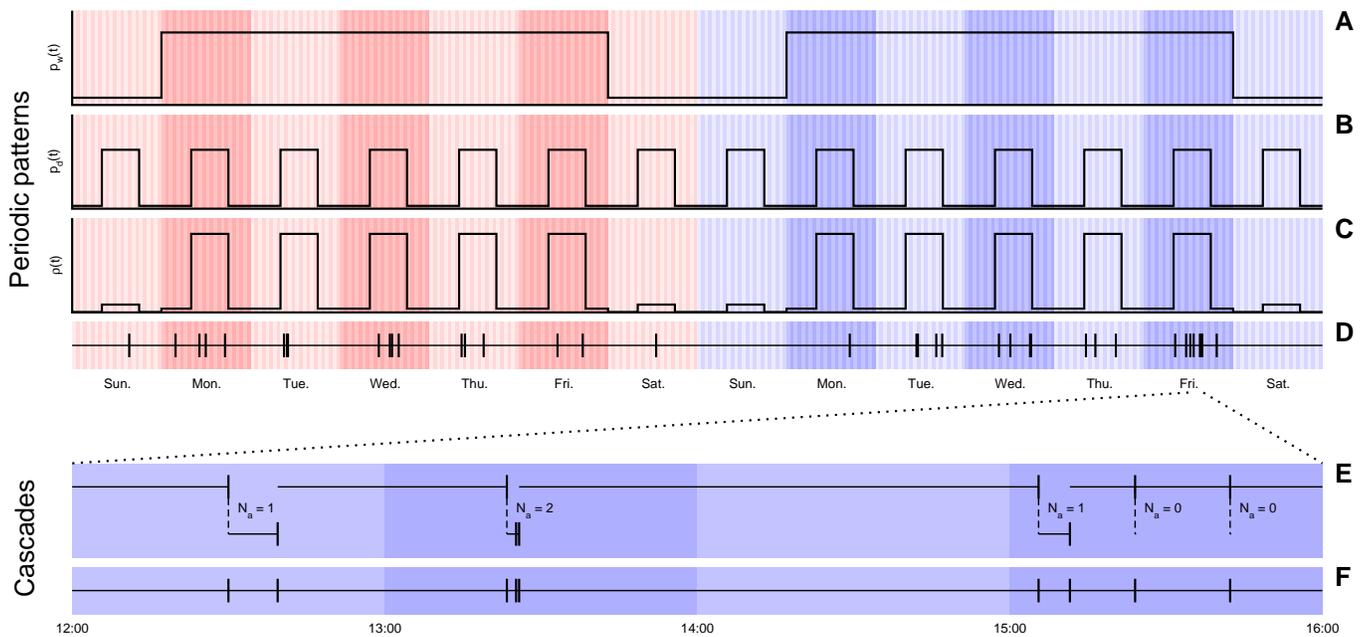}}
\caption{Example of a periodic and cascading stochastic process.
{\bf A}, Expected probability of starting an active interval during a
particular day of the week $p_w(t)$.  We depict two weeks to emphasize
that this pattern is periodic and that every week is statistically
identical to every other week.  We surmise that e-mail users are more
likely to send e-mails on the same days of the week, a consequence of
regular work schedules.
{\bf B}, Expected probability of starting an active interval during a
particular time of the day $p_d(t)$.  Again, we depict 14 days to
emphasize that this pattern is periodic and that every day is
statistically identical to every other day.  We surmise that e-mail
users are more likely to send e-mails during the same times of the
day, a consequence of circadian sleep patterns.
{\bf C}, The resulting activity rate $\rho(t)$ for the \nhppp.  The
activity rate $\rho(t)$ is proportional to the product of the daily
and weekly patterns of activity where the proportionality constant
$N_w$ is the average number of active intervals per week
(\eqnref{\ref{eqn:nhppp}}).
{\bf D}, A time series of events generated by a \nhppp.  Each event in
this time series initiates a cascade of additional events, an active
interval.
{\bf E}, Schematic illustration of cascading activity.  During
cascades---active intervals---we expect that an individual will send
$N_a$ additional e-mails according to a homogeneous Poisson process
with rate $\rho_a$.  We denote the start of active intervals with a
dashed line to signify that the activity is no longer governed by the
\nhppp~rate $\rho(t)$.  Once the active interval concludes, e-mail
usage is again governed by the periodic rate $\rho(t)$.  We refer to
the collection of active intervals as the active interval
configuration $\mathcal{C}$ throughout the manuscript.
{\bf F}, Observed time series.  Since the data does not isolate
intervals of activity, the observed time series is the superposition of
both the \nhppp~time series and the active interval time series.  
}
\label{fig:cnhppp_schematic}
\end{figure}

In order to gain some intuition about e-mail activity patterns, let us
consider a fictitious student, Katie.\footnote{We suspect that most
users only had access to their e-mail at the university since the data
are obtained from a European university prior to
2004~\cite{eckmann08}.}  Katie arrives at the university 20 minutes
before her Thursday morning class.  During this time, she decides to
check her e-mail and sends three e-mails.  Katie checks her e-mail
after lunch and sends a brief e-mail to a friend before her next
class.  Later that evening, Katie sends four more e-mails once she has
finished her homework.  Katie does not check her e-mail again until
the following day when she sends e-mails intermittently between
attending classes, completing homework assignments, and meeting social
engagements.  Katie spends the weekend without e-mail access and
doesn't send another e-mail until Monday.  Katie's e-mail activity,
which is similar to many e-mail users, is both \textit{periodic} and
\textit{cascading}.  That is, there are periodic changes in her
activity rate, which account for her sleep and work patterns, and
there are cascades of activity---active intervals---of varying length
when Katie primarily focuses on e-mail correspondence
(\figref{\ref{fig:cnhppp_schematic}}).

\begin{figure}
\centerline{\includegraphics*[width=1.0\columnwidth]{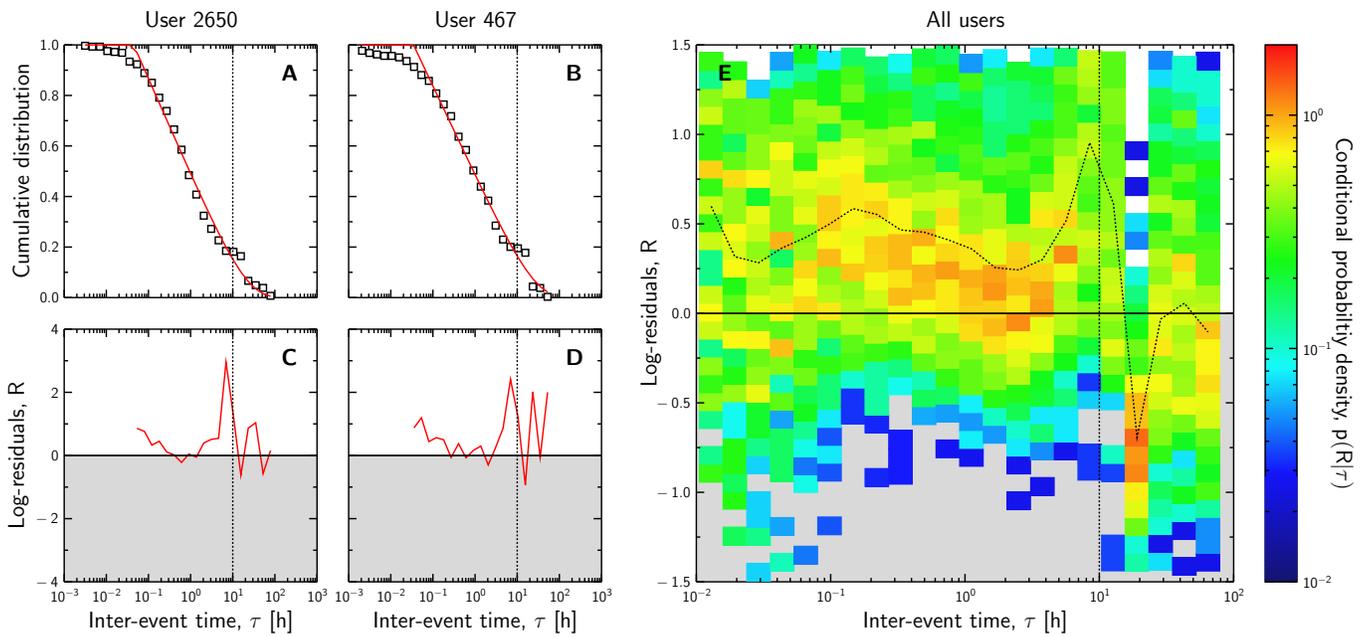}}
\caption{Systematic deviations of the data from the truncated
power-law null model due to periodic patterns of human activity.  The vertical 
lines at $\tau = 10$ hours is meant as a guide to the eye.
{\bf A}--{\bf B}, Comparison of truncated power-law model (red line)
with empirical data ($\square$) for Users 2650 and 467 from the data
set~\cite{eckmann04}.  Lines of best fit are estimated by minimizing
the area test statistic (\suppinfo{\secref{\tplref}}).
{\bf C}--{\bf D}, Log-residual, $R=\ln
(p_{\mathcal{M}}(\tau|\widehat{\theta})/p(\tau))$ of the best-fit
truncated power-law distribution model $\mathcal{M}$.  The shaded
region denotes inter-event times where the null model underestimates
the data.  If the empirical inter-event time distribution were
well-described by the truncated power-law null model, the
log-residuals $R$ would be small and normally distributed,
particularly in the tail of the distribution.  However, the
log-residuals $R$ have large systematic fluctuations in the tail of
the inter-event time distribution ($\tau>0.25$ hours) where the
power-law scaling approximately holds.
{\bf E}, Conditional probability density $p(R|\tau)$ obtained for all
\nusers~users under consideration.  The average log-residual at each
inter-event time is represented by the dashed line.  Both the average
log-residual and conditional probability density indicate that nearly
all users under consideration systematically deviate from the
truncated power-law null model, as anticipated from the arguments in
the main text.
}
\label{fig:pqm_residuals}
\end{figure}

If our intuition about deliberate human activity is correct, then the
periodic patterns of activity should manifest itself in the
inter-event time statistics, particularly when compared with the
predictions of the truncated power-law null model which does not
account for temporal periodicities (\suppinfo{\secref{\tplref}}).
Specifically, we anticipate that e-mail users typically send e-mails
during the same 8-hour periods of the day.  We therefore expect the
data to have significantly more inter-event times between $24 \pm 8$
hours---the time required to send e-mails on consecutive
workdays---than the truncated power-law model predictions.  We
therefore expect that the null model \textit{underestimates} the
number of inter-event times between 16 and 32 hours.
Due to the normalization of the probability density, the truncated
power-law model will over-estimate other inter-event times.
These predictions are all confirmed by the data, suggesting that
periodicity is a fundamental aspect of human activity
(\figref{\ref{fig:pqm_residuals}}).

\section{Model}

\begin{figure}
\centerline{\includegraphics*[width=1.0\columnwidth]{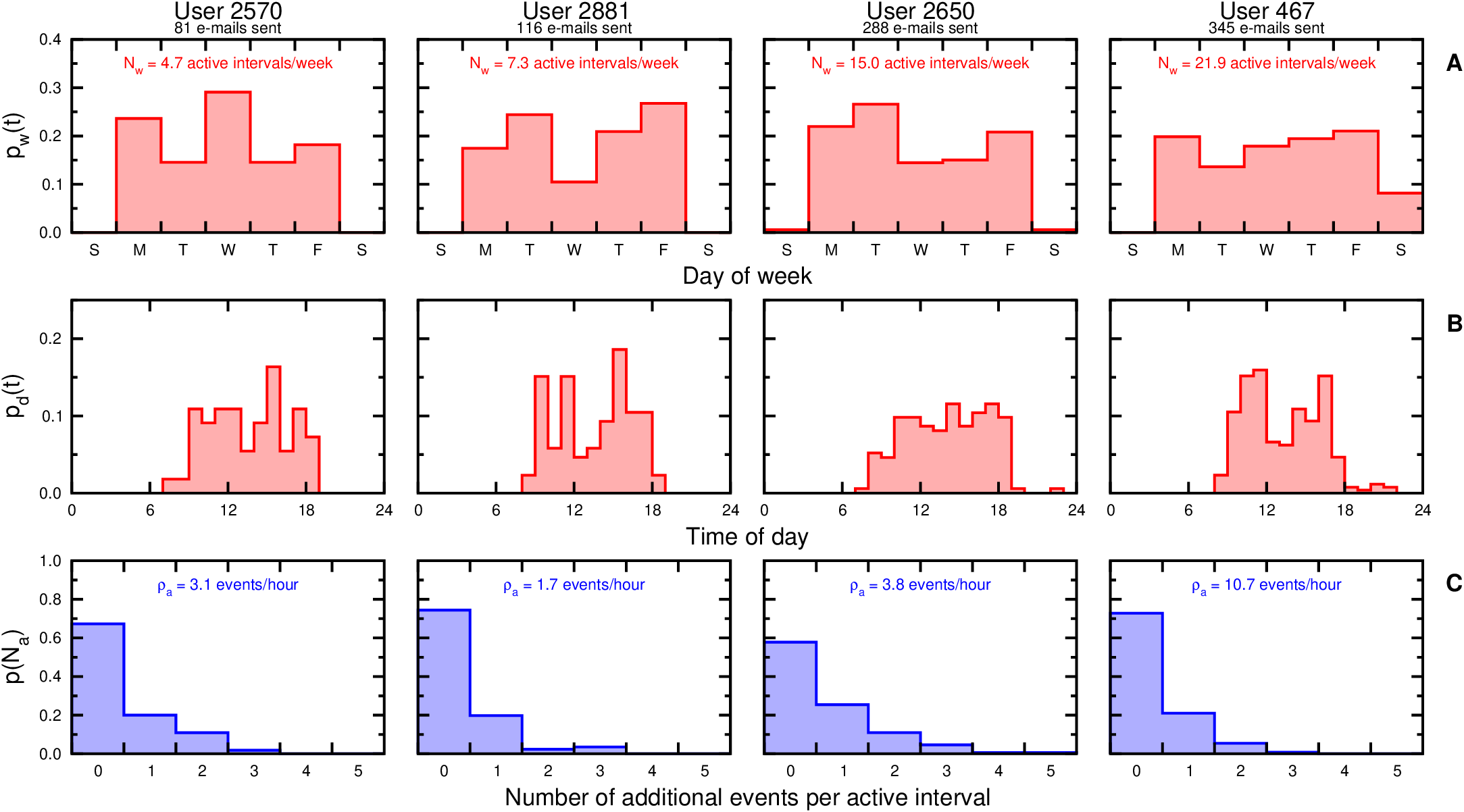}}
\caption{Patterns of e-mail activity for four users in increasing
order of e-mail usage (\suppinfo{\figref{\cnhpppfigref}} for the same
analysis for all \nusers~users).  These e-mail users exemplify the
e-mail usage patterns that are typical of the users in the data set.
We use simulated annealing to identify active intervals and calculate
the parameters for the \cnhppp~(\methods).  
%
%
The red distributions and text (A--B) correspond with the parameters
for the primary process, a \nhppp, while the blue distributions and
text (C) correspond with the parameters for the secondary process, a
homogeneous Poisson process.
{\bf A}--{\bf B}, Active intervals are much more likely during
weekdays rather than weekends and during the daytime rather than the
nighttime.  These prolonged periods of inactivity lead to the heavy
tail in the inter-event time distribution.
{\bf C}, Small inter-event times, in contrast, are characterized by
active intervals.  One can interpret active intervals in several ways:
larger $\rho_a$ may indicate that a user is a more proficient e-mail
user; larger $\langle N_a \rangle/\rho_a$ may suggest that an
individual has a larger attention span; $N_a/\rho_a$ may be the time
that an individual has to check e-mail before their next commitment.
}
\label{fig:activity_patterns}
\end{figure}

We propose a model of e-mail usage that incorporates the
hypothesized periodic and cascading features of human activity.  We
account for periodic activity with a primary process, which we model
as a \emph{non-homogeneous} Poisson process.  Whereas a homogeneous
Poisson process has a constant rate $\rho$, a \nhppp~has a rate
$\rho(t)$ that depends on time.  In our model, the rate $\rho(t)$
depends on time in a periodic manner; that is, $\rho(t) = \rho(t +
W)$, where $W$ is the period of the process.  Consistent with our
observations (\figref{\ref{fig:activity_patterns}}), we relate the
rate of the \nhppp~to the daily and weekly distributions of active
interval initiation, $p_{d}(t)$ and $p_{w}(t)$:
\begin{eqnarray}
\rho(t) & = & N_{w} p_{d}(t) p_{w}(t),
\label{eqn:nhppp}
\end{eqnarray}
where the period $W$ is one week and the proportionality constant
$N_w$ is the average number of active intervals per week.\footnote{In
specifying $N_w$ as the average number of active intervals per week,
we are implicitly assuming that the fraction of time spent in active
intervals is very small.  We have verified that this is the case for
all users under consideration.  Also, it is important to choose the
time step $\Delta t$ in the binning of the empirical $p_d(t)$ to be
sufficiently small such that the probability of an event occurring at
time $t$ is $\rho(t)\Delta t \ll 1$.  We choose $\Delta t = 1/N_w$
hours, which meets this criterion while still maintaining
computational feasibility.}

We further assume that each event generated from the primary process
initiates a secondary process, which we model as a homogeneous Poisson
process with rate $\rho_a$.  We refer to these ``cascades of
activity'' as \emph{active intervals}, during which $N_a$ additional
events occur where $N_a$ is drawn from some distribution $p(N_a)$.
Once the $N_a$ events have occurred in the active interval, the
activity of the individual is again governed by the primary process
defined by \eqnref{\ref{eqn:nhppp}}.  Our model thus mimics how
individuals like Katie use e-mail: Katie sends e-mails sporadically
throughout the day, but once she starts checking her e-mail, it is
relatively easy to send additional e-mails in rapid succession.  We
refer to the resulting model as a \cnhppp.\footnote{Our model is
similar in spirit to the Neyman-Scott cascading point
process~\cite{neyman58,lowen05} and the Hawkes self-exciting
process~\cite{hawkes71}, except that in our model (\emph{i}) the
primary process is modulated periodically by a 
non-homogeneous rate, and (\emph{ii})
the active intervals are non-overlapping.}

\section{Results}

To compare our model with the empirical data, we first need to
estimate the parameters of our model from the data.  Ideally, the data
would specify which events belong to the same active intervals---the
active interval configuration $\mathcal{C}$---so that we could
estimate the distributions $p_d(t)$, $p_w(t)$, and $p(N_a)$.  The data
we analyze, however, does not specify the actual active interval
configuration $\mathcal{C}_o$ so it is not evident whether, for
example, $p(N_a)$ should be described by a normal or exponential
distribution.

Because we do not know \emph{a priori\/} the functional form of the
activity pattern in the cascading process, we cannot use the formalism
implemented by, for example, Scott and co-workers \cite{scott99,scott03}.
Instead, we introduce a new method that enables us to nonparametrically infer the
empirical distributions $p_d(t)$, $p_w(t)$, and $p(N_a)$ from the
data.  

Given a particular active interval configuration $\mathcal{C}$, we can
easily calculate all of our model's parameters and compare it's
predictions with the empirical data:
$N_w$ is the average number of active intervals per week;
$p_d(t)$ and $p_w(t)$ are the probabilities of starting an active
interval at a particular time of day and week respectively;
the active interval rate $\rho_a$ is the inverse of the average
inter-event time in active intervals; and
the probability of $N_a$ additional events occurring during an active
interval $p(N_a)$ is estimated directly from the active interval
configuration (\figref{\ref{fig:activity_patterns}}).
We then manipulate the active interval configuration $\mathcal{C}$ to
find the active interval configuration $\widehat{\mathcal{C}}$ that
gives a best estimate of the observed inter-event time distribution
(\methods).  This method allows us to infer the best-estimate
distributions $\widehat{p}_d(t)$, $\widehat{p}_w(t)$, and
$\widehat{p}(N_a)$ given the data and our proposed model without
making any assumptions on their functional forms.

We next compare the predictions of the \cnhppp~with the empirical
cumulative distribution of inter-event times $P(\tau)$ for all
\nusers~users under consideration in the present study
(\suppinfo{\figref{\cnhpppfigref}}).  Since we are using the empirical
data to estimate the parameters for our model---that is, the estimated
parameters \emph{depend} on the data---we must use Monte Carlo
hypothesis testing~\cite{dagostino86,press02} to assess the
significance of the agreement between the predictions of our model and
the empirical data (\suppinfo{\secref{\mchtref}}).  The visual agreement of our model's
predictions are confirmed by $p$-values clearly above our 5\% rejection
threshold (\figref{\ref{fig:cnhppp_predictions}}).

\begin{figure}
\centerline{\includegraphics*[width=0.5\columnwidth]{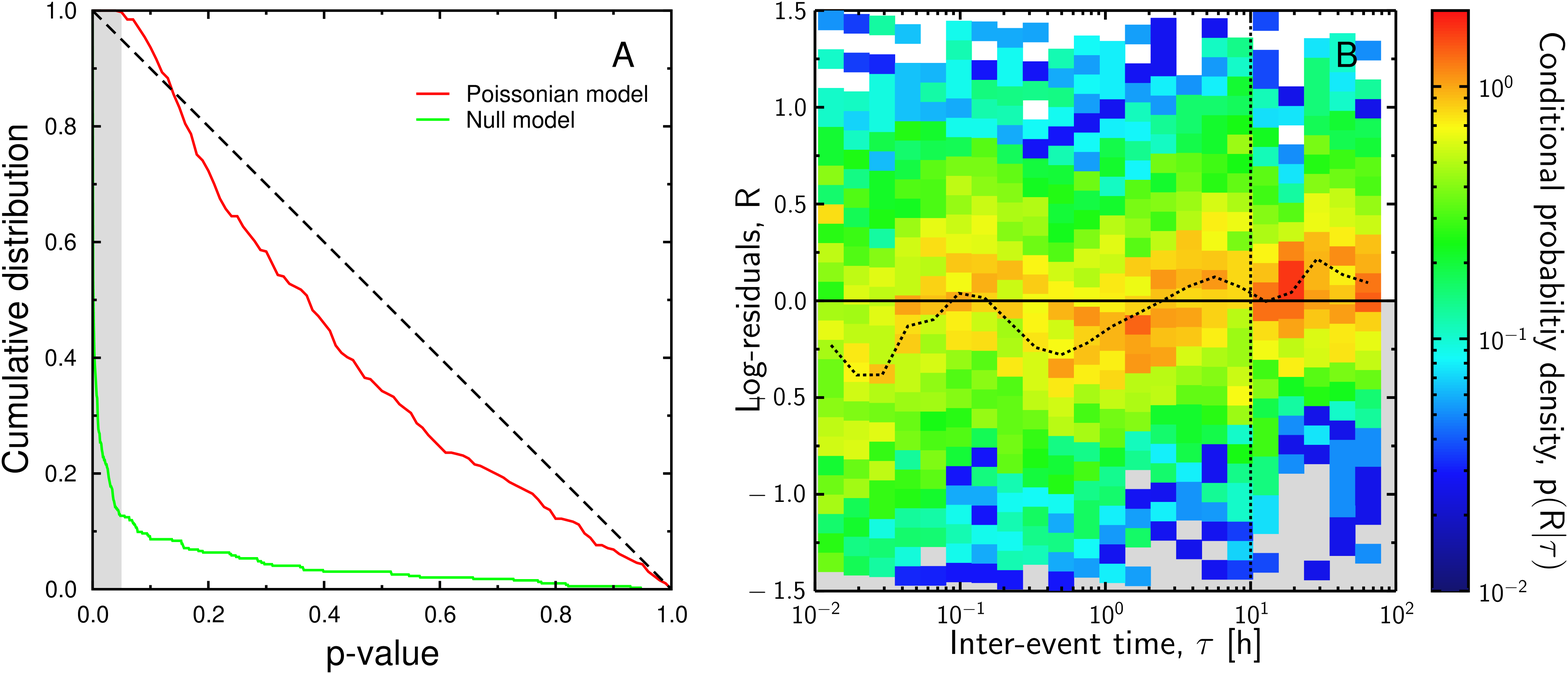}}
\caption{
{\bf A}, Summary of the hypothesis testing results for the \cnhppp~and the
truncated power-law null model for the \nusers~users under
consideration.
For each user, we compute the $p$-value between their inter-event time
distribution and the predictions of each model (\suppinfo{\secref{\mchtref}}).  We reject
a model for a particular user if the $p$-value is less than the 5\%
rejection threshold (gray shaded region).  At this significance level,
the \cnhppp~can be rejected for~\cnhpppnrejected~user while the
truncated power-law null model can be rejected for~\tplnrejected~users
(\suppinfo{\secref{\tplref}}).
Note that if the data were actually generated by one of the models
tested, we would expect to see a uniform distribution of $p$-values
(dashed line).  Since this is very nearly the case for the \cnhppp,
this provides additional evidence that our model is 
consistent with the data.
{\bf B}, Conditional probability density $p(R|\tau)$ obtained for all
\nusers~users under consideration.  The average log-residual at each
inter-event time is represented by the dashed line.  In contrast to the results
in~\figref{\ref{fig:pqm_residuals}E}, we find no systematic deviations between
the model predictions and the data in the tail of the inter-event time
distribution where the power-law scaling approximately holds.}
\label{fig:pval_comparison}
\end{figure}

In fact, the \cnhppp~can only be rejected at the 5\% significance level
for~\cnhpppnrejected~user, indicating that our model can not be rejected as a
model of human dynamics.  By comparison, the truncated power-law null model is
rejected at the 5\% significance level for~\tplnrejected~users.  Indeed, the
null model is always rejected for many more users than the \cnhppp~regardless
of the rejection threshold selected and our model displays none of the
systematic deviations from the data observed for the truncated power-law null
model (\figref{\ref{fig:pval_comparison}})

\section{Discussion}

Our results clearly demonstrate that circadian and weekly cycles, when coupled
to cascading activity, can accurately describe the heavy-tails observed in
email communication patterns. The question then is, would rational decision
making together with circadian and weekly cycles be equally able to describe
the statistical patterns observed for e-mail communication?  Even if the answer
to this question is affirmative, parsimony suggests that rational decision
making is not a necessary component of human activity patterns, given our
simpler explanation.

In addition to providing a good description of e-mail communication
patterns, we surmise that our model is readily applicable to many
other conscious human activities.  For instance, most people make
telephone calls sporadically throughout the day.  After a telephone
call has been made, it is effortless to make another telephone call.
Similarly, individuals run errands throughout the month.  Once an
individual runs one errand, it is easier to run another errand during
the same trip than it is to run errands again the following day.  Both
of these anecdotes are illustrative of the way humans tend to optimize
their time and effort to accomplish the tasks in their daily routines,
a process that is captured by the periodic and cascading mechanisms in
our model.

The particular periodic and cascading features that are incorporated
into our model depend on the activity under consideration.  For
instance, sexual activity is influenced by menstrual
cycles~\cite{udry68} and airline travel is influenced by
seasonality~\cite{kulendran97}.
Furthermore, our model can also be generalized to cases in which the
parameters are not stationary.  This may be important, for instance,
in the case of Darwin and Einstein's letter correspondence in which
the number of letters sent per year increases 100-fold over 40
years~\cite{oliveira05,vazquez06a}.

Although our model is only designed to account for a single activity
(e-mail correspondence), it can easily be extended to incorporate the
multitude of activities in which any individual participates.  To
facilitate the inclusion of additional activities, it is useful to
interpret our model as a non-stationary hidden Markov point
process~\cite{elliott95,scott99}.  Within this framework, an
individual switches between any two activities $i$ and $j$ with some
probability defined by a non-stationary Markov transition matrix
$T_{ij}(t)$ which depends on time $t$.  For instance, our model can be
redefined as a non-stationary hidden Markov point process which
switches between two states: a state in which an individual is not
composing e-mails and a state in which an individual is composing
e-mails.  Predictions of models that incorporate more than one
activity can then be verified against data that records several
activities for a single individual.

Our model further suggests a novel experiment~\cite{watts07} which not only
records when an individual has sent an e-mail, but also when that individual is
using a computer or actively utilizing an e-mail client.  This additional data
would provide direct empirical evidence for describing active intervals.  In
the absence of such data, we have developed a simulated annealing procedure
which allows us to nonparametrically infer the hidden Markov structure of our
model, providing insight into how to compare our model with other cascading
point processes~\cite{neyman58,lowen05}.

While our model provides an accurate description of \emph{when} an
e-mail is sent, a question left unaddressed is to determine
\emph{whom} the probable recipient of that e-mail is going to be.  For
instance, one might speculate that e-mails are sent randomly with some
Poissonian rate to acquaintances or individuals which share common
interests.  Alternatively, it is plausible that e-mails are sent based
on a perceived priority of important tasks, perhaps in response to
previous correspondence~\cite{barabasi05}.  When combined with our
model that statistically describes when individuals send e-mails,
quantifying the likely recipient of an e-mail will provide an
important step toward describing how the structure of e-mail and
social networks evolve.

Our study also provides a clear demonstration of how hypothesis
testing~\cite{dagostino86,sivia06} can objectively assess the validity
of a proposed model---a procedure we vehemently advocate.  Using this
methodology, we demonstrate that while both models reproduce the
asymptotic scaling of the observed inter-event time distribution, our
model is consistent with the entire inter-event time distribution
whereas the truncated power-law null model is not.

The consequences of our findings are clear; demonstrating that a model
reproduces the asymptotic power-law scaling of a distribution does not
necessarily provide evidence that the model is an accurate mechanistic
description of the underlying process.  Indeed, there is mounting
evidence that some purported power-law distributions in complex
systems may not be power-laws at
all~\cite{perline05,edwards07,clauset07}.  There may be a
common explanation for these apparent power-laws: complex systems are
inherently hierarchical but the distinct levels in the hierarchy are
difficult to distinguish~\cite{sales-pardo07}.  In the case of e-mail
correspondence for example, the active intervals are 
not recorded in the data
data, thereby concealing the various scales of e-mail activity.  This
demonstrates how the mixture of scales of activity can give rise to
scale-free activity patterns.  We suspect that similar
mixture-of-scales
explanations~\cite{silcock54,harris68,hausdorff96,willinger02,motter05}
may provide a basis for the reported universality of heavy-tailed
distributions in complex systems.

\section{Methods}

\subsection{Area test statistic}

We quantify the agreement between a model $\mathcal{M}(\theta)$ with parameters
$\theta$ and data set $\mathcal{D}$ by measuring the area $A$ between the
empirical cumulative distribution function $P_{\mathcal{D}}(u)$ and the model
cumulative distribution function $P_{\mathcal{M}}(u|\theta)$:
\begin{eqnarray}
A & = & \int \left| P_{\mathcal{D}}(u) - P_{\mathcal{M}}(u|\theta) \right| du .
\label{eqn:test_statistic}
\end{eqnarray}
We specify $u=\ln \tau$, which is roughly uniformly distributed, to
improve the numerical efficiency of our simulated annealing procedure.
The area test statistic is advantageous as it is easy to interpret and it
retains more information about the distribution than many other test
statistics~(\suppinfo{\secref{\teststatref}}).

\subsection{Identifying active intervals}

If we know the actual active interval configuration $\mathcal{C}_o$,
it would be straightforward to compute the parameters
$\theta_o=\{N_w,p_d(t),p_w(t),\rho_a,p(N_a)\}$ of the \cnhppp.  The
data, however, does not identify the actual active interval
configuration $\mathcal{C}_o$, we must use heuristic
methods~(\suppinfo{\secref{\simannealref}}) to determine the
best-estimate active interval configuration $\widehat{\mathcal{C}}$,
from which we can compute the best-estimate parameters
$\widehat{\theta}$.  We use simulated annealing to minimize the area test
statistic $A$ (\eqnref{\ref{eqn:test_statistic}}) for the inter-event
time distribution.  Thus, identifying active intervals that are
consistent with our expectations for our model reduces to finding the
best-estimate active interval configuration $\widehat{\mathcal{C}}$
which minimizes the area $A$ between the empirical data and
the predictions of the \cnhppp.

Our simulated annealing procedure is as follows.  Starting from a
random active interval configuration $\mathcal{C}$ in which adjacent
events are randomly assigned to the same active interval, we compute
the parameters $\theta$ of the \cnhppp, then we numerically estimate
the cumulative distribution $P_{\mathcal{M}}(u|\theta)$, and finally
we measure the area test statistic $A(\mathcal{C})$ of the active interval
configuration $\mathcal{C}$.  The active interval configuration is
modified to a new configuration $\mathcal{C}^{\prime}$ by either
merging two adjacent active intervals or by splitting an active
interval.  If the new configuration $\mathcal{C}^{\prime}$ reduces the
area test statistic, then the new configuration is unconditionally
accepted.  Otherwise the configuration is conditionally accepted with
probability $\exp(-(A(\mathcal{C}^{\prime}) - A(\mathcal{C}))/T)$,
where $T$ is the effective ``temperature'' measured in units of the
area test statistic $A$.  After attempting $2N$ configurations at each
temperature so that each pair of $N$ consecutive events might be
merged and split, we reduce the temperature $T$ by 5\% until the
active interval configuration settles at the best-estimate
$\widehat{\mathcal{C}}$ without moving for 5 consecutive cooling
stages.\footnote{Throughout the simulated annealing procedure, we track
the lowest area test statistic configuration.  If the system has settled
in a configuration which is not the lowest area test statistic
configuration, the system is placed in the lowest area test statistic
configuration and the system is cooled further.}  We have verified that our
simulated annealing procedure accurately identifies active intervals
and estimates parameters $\theta$ in synthetically generated
\cnhppp~data
sets~(\suppinfo{\secref{\simannealref}~and~\figref{\simannealfigref}}).

\begin{acknowledgments}
We thank R.~Guimer\`a, M.~Sales-Pardo, M.J.~Stringer,
E.N.~Sawardecker, S.M.~Seaver, and P.~McMullen for insightful comments
and suggestions.  R.D.M.~and D.B.S.~thank the NSF-IGERT program
(DGE-9987577) for partial funding during this project.  A.E.M. is
supported by the NSF under grant DMS-0709212.  L.A.N.A. gratefully
acknowledges the support of NSF award SBE 0624318 and of the
W.~M.~Keck Foundation.
\end{acknowledgments}


\end{article}

\clearpage

\clearpage

\clearpage

\clearpage

\begin{figure}
\centerline{\includegraphics*[width=1.0\columnwidth]{Figures/CNHPPP_Predictions/cnhppp_predictions}}
\caption{Comparison of the predictions of the \cnhppp~(red line) with
the empirical cumulative distribution of inter-event times $P(\tau)$
(black line) for the same users from
\figref{\ref{fig:activity_patterns}}
(\suppinfo{\figref{\cnhpppfigref}} for the same analysis for all
\nusers~users).
We use the area test statistic $A$ (\eqnref{\ref{eqn:test_statistic}}) and
Monte Carlo hypothesis testing to calculate the $p$-value between the
model and the data (\suppinfo{\secref{\mchtref}}).  As these figures are presented, the
area test statistic $A$ is the area between the two curves.
Not only do the predictions of the \cnhppp~visually agree with the
empirical data, but the $p$-values indicate that it can not be
rejected as a model of e-mail activity at a conservative 5\%
significance level.}
\label{fig:cnhppp_predictions}
\end{figure}

\clearpage

\end{document}